\documentclass{ws-procs9x6-notrim}
\usepackage{bm}
\usepackage{epsfig}
\usepackage{graphicx}
\usepackage{fancybox}
\def\centeron#1#2{{\setbox0=\hbox{#1}\setbox1=\hbox{#2}\ifdim
\wd1>\wd0\kern.5\wd1\kern-.5\wd0\fi
\copy0\kern-.5\wd0\kern-.5\wd1\copy1\ifdim\wd0>\wd1
\kern.5\wd0\kern-.5\wd1\fi}}
\def\centerover#1#2{\centeron{#1}{\setbox0=\hbox{#1}\setbox
1=\hbox{#2}\raise\ht0\hbox{\raise\dp1\hbox{\copy1}}}}
\def\centerunder#1#2{\centeron{#1}{\setbox0=\hbox{#1}\setbox
1=\hbox{#2}\lower\dp0\hbox{\lower\ht1\hbox{\copy1}}}}
\def\lsim{\;\centeron{\raise.35ex\hbox{$<$}}{\lower.65ex\hbox
{$\sim$}}\;}
\def\gsim{\;\centeron{\raise.35ex\hbox{$>$}}{\lower.65ex\hbox
{$\sim$}}\;}
\begin{document}

\rightline{\vbox{\halign{&#\hfil\cr
&\normalsize ANL-HEP-CP-02-114\cr
%&\normalsize UIC-87-14\cr
%&\normalsize\today\cr
}}}
\vspace{-3 ex}

\title{Recent Results on Quarkonium Production and Decay}

\author{Geoffrey T.~Bodwin}

\address{High Energy Physics Division, \\
Argonne National Laboratory, \\ 
Argonne, IL 60439, USA\\ 
%E-mail: gtb@hep.anl.gov
}


\maketitle

\abstracts{I summarize the current status of the comparison between
experiment and the predictions of the NRQCD factorization approach to
quarkonium decay and production. I also present the results of some
recent calculations and theoretical developments in the NRQCD
factorization approach.}

%%%%Start of Text%%%%%%%%%%%%%%%%%%%%%%%%%%%%%%%%%%%%%%%%%%%%%%%%%%%%%%%%%%%%

\section{Some Successes of the NRQCD Factorization Approach}

Since its formulation, the NRQCD factorization approach\cite{bbl} has 
enjoyed a number of successes in predicting inclusive decay rates and 
production cross sections for quarkonium states.

The initial success of the NRQCD factorization method was in the
computation of infrared-finite predictions for the inclusive decay rates
of $P$-wave quarkonium states.\cite{bbl-decays} A subsequent global fit
to the $P$-wave charmonium data\cite{maltoni} has yielded a $\chi^2$
per degree of freedom of $15.0/10$ and values of the nonperturbative
NRQCD matrix elements that are in good agreement with lattice
determinations\cite{bks-charmonium} and with estimates based on the NRQCD
velocity-scaling rules.\cite{bbl,petrelli-et-al}

In the area of quarkonium production, the first success of the NRQCD
factorization approach was in explaining the Tevatron data for the
inclusive production of $J/\psi$, $\chi_c$, $\psi(2S)$, $\Upsilon$, and
$\Upsilon(2S)$ states. Previous calculations in the color-singlet model
had yielded results that were smaller than the data by more than an
order of magnitude. In the NRQCD predictions, the unknown NRQCD matrix
elements were determined from fits to the data. Nevertheless, the
comparison with the Tevatron data gives a nontrivial confirmation of the
NRQCD factorization approach because the extracted matrix elements
satisfy the velocity-scaling rules, and the shape of the data as a
function of $p_T$ is consistent with NRQCD factorization, but not with
the color-singlet model.
%\begin{figure}[ht]
%\centerline{\includegraphics[height=7.0cm]{kramer-fig2.eps}}
%\caption{CDF $J/\psi$ production}
%\end{figure}

A recent success of the NRQCD factorization approach is in predicting the 
cross section for $\gamma \gamma\rightarrow J/\psi +X$ at 
LEP.\cite{klasen-kniehl-mihaila-steinhauser} In this case, the prediction 
makes use of NRQCD matrix elements extracted from the Tevatron data, and 
hence, provides a test of the predicted universality of the matrix 
elements. The principal theoretical uncertainties arise from 
uncertainties in the renormalization and factorization
scales and from estimates of the color-octet matrix
elements. The data from the Delphi experiment are consistent with the 
NRQCD factorization prediction and clearly disfavor the prediction of 
the color-singlet model.
%\begin{figure}[ht]
%\centerline{\epsfig{figure=delphi1.eps,width=10cm}}
%\caption{Delphi}
%\end{figure}

Another prediction that makes use of the NRQCD matrix elements extracted
from the Tevatron data is that for quarkonium production in
deep-inelastic scattering at HERA.\cite{kniehl-zwirner} Again, the
theoretical uncertainties arise mainly from uncertainties in the
renormalization and factorization scales and in the color-octet matrix
elements. Data from the H1 experiment plotted as
functions of $p_T$ or $Q^2$ favor the prediction of NRQCD factorization
over that of the color-singlet model. Neither prediction is entirely
consistent with the data as a function of $z$. It should be noted that
the most recent calculation\cite{kniehl-zwirner} disagrees with a number
of previous theoretical results, and that these disagreements have not
yet been resolved fully.

\section{Some Problematic Comparisons with Experiment}

There are some processes for which the comparisons between the 
experimental data and the NRQCD factorization predictions are less 
satisfactory.

The production of transversely polarized quarkonium at the
Tevatron is potentially a ``smoking gun'' for the color-octet production
mechanism, which is an integral part of the NRQCD factorization
formalism. For quarkonium production at large $p_T$ ($p_T\gsim 4m_c$ for
$J/\psi$), gluon fragmentation into quarkonium via the color-octet
mechanism is the dominant process. At large $p_T$, the fragmenting
gluon is nearly on mass shell, and, so, is transversely polarized. NRQCD 
predicts\cite{cho-wise} that the polarization is largely transferred to the 
$J/\psi$, although it is diluted somewhat by non-fragmentation 
processes, radiative corrections, and feeddown from higher quarkonium 
states.\cite{beneke-rothstein-pol,beneke-kramer-pol,braaten-kniehl-lee}
The data from the CDF experiment show
no evidence for the predicted increase of transverse polarization with
increasing $p_T$. However, the error bars are large,
and only the highest-$p_T$ data point is actually inconsistent with the
prediction. There are also large theoretical uncertainties, primarily 
from corrections of higher order in $\alpha$ and $v$, including 
$v^2$ effects in the polarization transfer.

Another problematic process is inelastic quarkonium photoproduction at
HERA. The HERA data\cite{H1-photoproduction,Zeus-photoproduction} are just 
barely compatible with the 
prediction.\cite{cacciari-kramer,amundson-fleming-maksymyk,ko-lee-song,%
kniehl-kramer-photo}
The color-octet mechanism leads to a prediction of
increasing rate with increasing energy fraction $z$. This is not
observed. However, theoretical uncertainties arising from corrections of
higher order in $\alpha_s$ and $v$, uncertainties in $m_c$, and the
breakdown of the $v$ expansion near $z=1$ have not been addressed or
have been addressed 
incompletely.\cite{kramer-color-singlet-hera,beneke-rothstein-wise,%
beneke-schuler-wolf}
It should be noted that the experimental data differential in $p_T$ are 
compatible with color-singlet production
alone,\cite{kramer-color-singlet-hera} even at large $p_T$. 

A recent striking result from the Belle collaboration\cite{belle} concerns 
production of a double $c\bar c$ pair in $e^+e^-$ collisions:
$\sigma(e^+e^-\rightarrow J/\psi\,c\bar c)
/\sigma(e^+e^-\rightarrow J/\psi\, X)=0.59^{+0.15}_{-0.13}\pm 0.12$. 
Perturbative QCD plus the color-singlet model leads to the 
prediction\cite{cho-leibovich}
$\sigma(e^+e^-\rightarrow J/\psi\,c\bar c)/
\sigma(e^+e^-\rightarrow J/\psi\, X)\approx 0.1$.
This seems to be a major discrepancy between theory and experiment. 

\section{General Difficulties in the Theory}

There are several difficulties that arise when one attempts to make accurate
theoretical predictions for quarkonium production and decay processes.

Foremost among these is the fact that the color-octet NRQCD matrix
elements are poorly determined. For some processes, only linear
combinations of color-octet matrix elements can be determined from the
data with reasonable accuracy. Different linear combinations are fixed
by different processes---a situation that makes it difficult to test the
universality of the color-octet matrix elements.

A further difficulty in making accurate theoretical predictions is the
fact that corrections to the short-distance coefficients of
next-to-leading-order (NLO) in $\alpha_s$ are often large. For example,
for the process $J/\psi\rightarrow \gamma\gamma\gamma$, the NLO
correction is $-12.62\alpha_s/\pi$, relative to the leading-order
coefficient. A related problem is that the dependence of the calculated
short-distance coefficients on the renormalization scale is often large.
These issues raise doubts about the convergence of the perturbation
expansion for the short-distance coefficients. For some key processes,
such as $J/\psi$ production in hadronic collisions, the NLO corrections
have not yet been computed. For certain processes, resummations of large
logarithms of $p_T^2/M_\psi^2$, $M_Z^2/M_\psi^2$, and $z^2$ have been
carried
out.\cite{braaten-doncheski-fleming-mangano,boyd-leibovich-rothstein} In 
some instances, resummations of logarithms of $1-x$ and $x$ may also be 
required.

Another significant source of inaccuracy in theoretical predictions is 
the existence of large relativistic corrections. For example, 
for the process $J/\psi\rightarrow \gamma\gamma\gamma$, the order-$v^2$ 
correction is $-5.32 v^2$. Since $v^2\approx 0.3$ for charmonium and 
$v^2\approx 0.1$ for bottomonium, one can question whether the $v$ 
expansion converges. Furthermore, for many processes, the order-$v^2$ 
corrections have not yet been computed.

In the remainder of this paper, I describe some recent progress in 
confronting these obstacles to accurate theoretical predictions.

\section{Relativistic Corrections to Gluon Fragmentation into
$\bm{S}$-wave Quarkonium and to $\bm{S}$-wave Quarkonium Decay}

Motivated by the importance of the fragmentation process for $S$-wave
quarkonium production at large $p_T$, J.~Lee and I calculated the
order-$v^2$ corrections to that process. Our preliminary results are
that the correction to the color-singlet process is approximately 74\%
for the $J/\psi$, and the correction to the color-octet process is
approximately -54\% for the $J/\psi$.  The correction to the
color-singlet process reduces the predicted value of the transverse
polarization parameter $\alpha$ by about 10\% at large $p_T$. The
corrected prediction is still far above the highest-$p_T$ CDF data
point. The correction to the color-octet process directly affects the
value of the ${}^3S_1$ color-octet matrix element that is obtained by
fitting to the Tevatron data.

A.~Petrelli and I have computed the coefficients of color-singlet
operators of order $v^4$ for S-wave quarkonium decays.\cite{bodwin-petrelli} 
This is the first decay or production calculation
at next-to-next-to-leading order in $v$. The series for ${}^1S_0$ decay
into two photons and for the color-singlet part of ${}^1S_0$ decay
into light hadrons are $1-1.33v^2+1.51v^4+\cdots$, the series for
${}^3S_1$ decay into $e^+e^-$ is $1-1.33v^2+1.61v^4+\cdots$, and the
series for the color-singlet part of ${}^3S_1$ decay into light hadrons
is $1-5.32v^2+7.62v^4+\cdots$. All of these series seem to be converging
in order $v^4$, even the one for ${}^3S_1$ decay into light hadrons,
which receives a large contribution in order $v^2$. It is known that
color-octet contributions in the first non-trivial order ($v^3$ and
$v^4$) are large.\cite{maltoni,bodwin-petrelli} The hope is that these, too, 
will receive small corrections in the next order in $v$.

\section{Lattice Computation of Bottomonium Decay Matrix Elements}

D.K.~Sinclair, S.~Kim, and I have recently computed matrix elements for
bottomonium decay on the lattice, using two dynamical light
quarks.\cite{bks-bottomonium} The essential steps in this computation
are to measure matrix elements in a lattice simulation, to use
perturbation theory (at the cutoff scale $m_b$) to relate the lattice
and continuum matrix elements, and to extrapolate to three light-quark
flavors and to physical light-quark masses. For the matrix element of
the color-singlet $S$-wave operator in the $\Upsilon$ state, we obtain
$4.10(1)(9)(41) \hbox{ GeV}^3$, which compares well with the
phenomenological value $3.86(14) \hbox{ GeV}^3$ extracted from the
measured rate for $\Upsilon\rightarrow e^+e^-$. We also obtained
values for matrix elements of the
color-singlet $P$-wave operator and the color-octet $S$-wave operator in
the $\chi_b$ state, which can be used to make predictions for the decays
of $\chi_b$ states. The matrix element of the order-$v^2$ color-singlet
$S$-wave operator in the $\Upsilon$ state is poorly determined, owing to
large lattice-to-continuum corrections. The values of all of these matrix
elements are strongly affected by the inclusion of light dynamical
quarks in the calculation.

\section{Resummation of QCD Corrections to Quarkonium Decay Rates}

%\subsection{$\bm{\eta_c}$ Decay in NLO}

%$$
%\Gamma(\eta_c\rightarrow gg)=
%{\langle\eta_c|\chi^\dagger\psi\psi^\dagger\chi|\eta_c\rangle\over m_c^2}
%{\pi C_F\over 2 
%N_c}\alpha_s^2(\mu)
%\biggl[1+a_{gg}^{(1)}{\alpha_s(\mu)\over\pi}+{4\over 3}{\langle p^2\rangle
%\over m_c^2}+O(\alpha_s^2)+O(v^3)\biggr ].
%$$
%
%$$
%\Gamma(\eta_c\rightarrow \gamma\gamma)
%={\langle\eta_c|\chi^\dagger\psi\psi^\dagger\chi|\eta_c\rangle\over m_c^2}
%2\pi Q^4\alpha_{em}^2
%\biggl[1+a_{\gamma\gamma}^{(1)}{\alpha_s(\mu)\over\pi}+{4\over 3}{\langle 
%p^2\rangle
%\over m_c^2}+O(\alpha_s^2)+O(v^3)\biggr].
%$$
%
%$$
%\langle p^2\rangle={\langle \eta_c|\upon 1/2
%[\chi^\dagger (\upon i/2 \tensor{\bf D})^2
%\psi+\hbox{H.c.} |\eta_c\rangle\over
%\langle \eta_c|\chi^\dagger \psi\psi^\dagger\chi |\eta_c\rangle}.
%$$
In this Section I describe work carried out in collaboration with Y.-Q. 
Chen.\cite{bodwin-chen}

Expressions for the decay rates of the $\eta_c$ into light hadrons (two
gluons in perturbation theory) and two photons are both known through
NLO in $\alpha_s$ and $v$. Each expression depends on
two NRQCD matrix elements. However, if we take the ratio of the rates,
the dependence on the matrix elements cancels:
\begin{eqnarray}
R^{\rm NLO}(\mu)&=&{\Gamma(\eta_c\rightarrow gg)
\over \Gamma(\eta_c\rightarrow \gamma\gamma)}
=R_0 \biggl\{1+\biggl[\biggl({199\over 6}-{13\pi^2\over 8}\biggr)
-{8\over 9}n_f\biggr]
{\alpha_s(\mu)\over \pi}\nonumber\\
&&+2\beta_0\alpha_s\ln {\mu^2\over 
4m_c^2}+O(\alpha_s^2)+O(v^3)\biggr\},
\label{ratio}
\end{eqnarray}
where $R_0=9\alpha_s^2(\mu)/(8 \alpha_{em}^2)$, and 
$\beta_0\,=\, (33-2n_f)/(6\pi)$. 

One might hope that Eq.~(\ref{ratio}) would yield an accurate prediction
for the ratio. Unfortunately, the term in Eq.~(\ref{ratio}) proportional
to $\alpha_s$ is approximately $1.1R_0$,
casting doubt on the convergence of the perturbation expansion. Ignoring
this difficulty and setting $\mu=2m_c$, we obtain $R^{\rm
NLO}(2m_c)=2.1\times 10^3.$ The experimental value is $R^{\rm
Exp}=(3.3\pm 1.3)\times 10^3$. The agreement is reasonable, however the
$\mu$ dependence of the theoretical result is strong. For example, using
BLM scale setting\cite{BLM} we obtain $\mu_{\rm BLM}\approx 0.52m_c$.
This leads to an NLO term that is $0.5$ times the leading term, and, so,
the convergence appears to be better. However, the agreement with
experiment is poor: $R^{\rm NLO}(\mu_{\rm BLM})=9.9\times 10^3 $.

%\subsection{Resummation of Vacuum-Polarization Bubbles}

It is instructive to re-write Eq.~(\ref{ratio}) in terms of $\beta_0$:
\begin{eqnarray}
R^{\rm NLO}(\mu)&=&R_0
\biggl\{1+\biggl[\biggl({37\over 2 }-{13\pi^2 \over 
8}\biggr)
+\pi\beta_0\left({16\over 3}+2\ln {\mu^2\over 4 
m_c^2}\right)
\biggr]{\alpha_s(\mu) \over \pi}\biggr\}.\nonumber\\
\end{eqnarray}
The terms proportional to $\alpha_s$ that do not contain a factor
$\beta_0$ have a value $2.5R_0$, while the terms proportional to $\beta_0$
have a value $12R_0$ for $\mu^2=4m_c^2$. The fact that the terms
proportional to $\beta_0$ dominate in the NLO correction suggests that
we should resum the contributions proportional to $(\alpha_s\beta_0)^n$
to all orders in $\alpha_s$. To do this, we make use of the method of
naive non-Abelianization (NNA).\cite{beneke-braun} That is, we sum the
fermion-loop vacuum-polarization contributions and then take gluon loops
into account by replacing the fermion-loop contribution to $\beta_0$
with the full $\beta_0$.

In order to carry this out, one must compute all cuts through 
the two final-state complete gluon propagators, where each propagator 
consists of a chain of free gluon propagators and vacuum-polarization 
bubbles. The result is 
\begin{eqnarray}
R^{\rm Bub}=&&\sum_{n,m}\biggl[\int_0^1 {dx\over 2\pi x}
\int_0^1 {dy\over 2\pi y} 
f(x,y)I_R^{(n)}(x)I_R^{(m)}(y)\;\theta(1-\sqrt{x}-\sqrt{y})
\nonumber\\
&&+2G_{V}^{(n)}\int_0^1 {dx\over 2\pi x} f(x,0)I_R^{(m)}(x)
+f(0,0)\,G_{V}^{(n)} G_{V}^{(m)}\biggr],
\label{cuts}
\end{eqnarray}
where $x=k^2/(4m_c^2)$, $y=l^2/(4m_c^2)$, $k$ and $l$ are the gluon momenta, 
$f(x,y)$ is the phase-space and heavy-quark factor ($f(0,0)=1$),
%
%$$
%f(x,y)={[1-2(x+y)+(x-y)^2]^{3/2}\over (1-x-y)^2};
%$$
%$$
%f(0,0)=1.
%$$
%
$I_{R}^{(n)}(x)=-2\,{\rm Im}\,[-i\Pi(x)]^n$ is the sum of quark cuts of
the complete gluon propagator in $n$th order, $G_{V}^{(n)}=[-i\Pi(0)]^n$
is the sum of gluon cuts of the complete gluon propagator in $n$th
order, and $\Pi_{\mu\nu}(x)=(k^2g_{\mu\nu}-k_\mu k_\nu)\Pi(x)$ is the
one-loop vacuum polarization. In Eq.~(\ref{cuts}), the first, second, 
and third terms correspond to the $qq$, $qg$, and $gg$ cuts, 
respectively.

The individual terms in Eq.~(\ref{cuts}) are infrared divergent, but we can
write $R^{\rm Bub}=G_1^2+G_2$, where 
\begin{equation}
G_1=\sum_n \int_0^1 {dx\over 2\pi x}\, f(x,0)I_R^{(n)}(x)+\sum_n 
G_{V}^{(n)},
\end{equation}
and
\begin{eqnarray}
G_2\equiv \sum_{n,m}\biggl[&&\int_0^1 {dx\over 2\pi x} \int_0^1
{dy\over 2\pi y} f(x,y)I_R(x)^{(n)} I_R^{(m)}(y)\;\theta(1-\sqrt{x}-\sqrt{y})
\nonumber\\
&&-\int_0^1 {dx\over 2\pi x} f(x,0) I_R^{(n)}(x)
\int_0^1 {dy\over 2\pi y} f(0,y) I_R^{(m)}(y)\biggr].
\label{g2}
\end{eqnarray}
$G_1$ is infrared finite because of the KLN theorem, and $G_2$ is 
infrared finite because the integrands in its two terms become identical 
when $x$ or $y$ vanishes.

%\subsection{Convergence of the Single Bubble Chain}

Let us first consider the properties of a single vacuum-polarization-bubble 
chain, as 
manifested in $G_1$. A calculation of $G_1$ yields 
\begin{eqnarray}
G_1(\mu)=&&{1 \over \pi \alpha_s(\mu)\beta_0 } \;
 \arctan { { \pi \alpha_s(\mu)\beta_0}\over
 {1-\alpha_s(\mu)\,\beta_0 \,d}}\nonumber\\
 &&\;-\; {\frac{1}\pi } \; \sum_{n=1}^\infty \;
 \int_{0}^1 \; d x \;{\rm Im\;}
 [\,\alpha _s(\mu )\,\beta_0\,
   \left(\,d-\ln x + i\pi \,\right)]^n,
\label{G1}
\end{eqnarray}
where $d$ is a renormalization-scheme-dependent constant.
Now, integration of $\ln^n x$ down to $x=0$ produces factorial growth:
$\int_0^1 dx\, x^m\ln^n x=(-1)^n n!/(m+1)^{n+1}$. Therefore, the second
term of $G_1$ may contain contributions that grow as $n!$. This is an
indication that, owing to the failure of asymptotic freedom at small
$x$, the perturbation series may diverge. Factorial growth in the
perturbation series is characteristic of the presence of a
renormalon---a singularity in the Borel transform of the decay
rate---that signals the importance of nonperturbative effects.

However, when the imaginary part of Eq.~(\ref{G1}) is computed, a 
seemingly miraculous cancellation of the factorial growth occurs.
The singularity in the Borel transform also vanishes in the 
imaginary part. In order to understand why this happens, let us 
re-write the second term of $G_1$ as a contour integral:
\begin{equation}
\sum_n\int_0^1 {dx\over \pi}\,{\rm Im}\,[\alpha_s\beta_0(-\ln x+d+i\pi)]^n
=\sum_n\int_C {dx\over 2\pi}\,[\alpha_s\beta_0(-\ln x+d)]^n,
\label{contour}
\end{equation}
where the contour runs from $-1$ to $0$ to $1$ around the cut in $\ln
x$. We can deform the contour out of the region of small $x$ into a
circle of radius unity. Now the entire contour lies in a region in which
perturbation theory applies. No factorial growth occurs. The series is
bounded by a geometric series and, hence, is convergent.

%\subsection{The Order of Limits Matters}

Suppose that, for the second term of $G_1$, we carry out the perturbation 
summation before performing the integration. Then, summing the resulting 
geometric series, we obtain 
\begin{equation}
%\int_0^1 {dx\over \pi}\,\sum_n {\rm Im}\,[\alpha_s\beta_0(-\ln 
%x+d+i\pi)]^n=
\int_0^1{dx\over\pi}\, {\rm Im}\biggl[{1\over 1-\alpha_s\beta_0(-\ln 
x+d+i\pi)}\biggr].
\label{opposite-order}
\end{equation}
This differs from the expression (\ref{contour}) by a small, but
nonzero, amount: $
[1/(\alpha_s\beta_0)]\exp\{-1/[\alpha_s(\mu)\beta_0]+d\} \approx
e^d{\Lambda_{\rm QCD}^2/(\alpha_s\beta_0}\mu^2)$,
which has the typical form of a nonperturbative contribution. Both 
orders of operations lead to convergent expressions. However,
$\log x$ becomes unbounded at small $x$, and, so, the convergence is not 
uniform, and different orders of limits can produce different results.
If we carry out the summation before performing the integration, as in 
the expression (\ref{opposite-order}), there is
a Landau pole at in the integrand at $x_0=\exp[-1/(\alpha_s\beta_0)+d]$.
The residue at this pole is the difference between the expressions with
different orders of operations.

This raises a question as to which order of limits, if either, is
correct. In the unintegrated expression, for $x$ sufficiently small, we
are outside the radius of convergence of the perturbation series, and
the perturbation sum is unreliable.  In the integrated expression, at
any finite order in $\alpha_s$, we can deform the $x$ contour to get out
of the region of small $x$.  Then, the perturbation expansion converges,
and we can take the $n\rightarrow \infty$ limit. Hence, this latter
order of limits is the only one of the two that leads to a perturbation
expansion that is not obviously unreliable.

%\subsection{The Nonperturbative Region in a Double Chain}

Now let us consider the properties of two vacuum-polarization-bubble 
chains, as manifested in 
$G_2$ [Eq.~(\ref{g2})]. For $x\sim 1$, kinematics force $y$ to be in the 
nonperturbative region near 0. 
But, $x\sim 1$ implies that $\sqrt{k^2}$ 
is large compared with $l$ or with the three-momentum of the heavy quark.
Then, in NRQCD, the propagator carrying $k$ can be shrunk to a point. 
The result is a one-loop correction to the color-octet operator
$\mathcal{O}_8({}^3S_1)$.
% = \psi^\dagger \mbox{\boldmath $\sigma$} T^a \chi
%\cdot \chi^\dagger \mbox{\boldmath $\sigma$} T^a \psi$. 
We subtract this 
nonperturbative piece from the perturbative calculation and absorb it into 
the quarkonium matrix element of $\mathcal{O}_8({}^3S_1)$.
After these subtractions, the $x$ and $y$ contours can be deformed out
of the nonperturbative region. Note that the contour-deformation
argument is crucial. Without it, there would be contributions from the
nonperturbative region in which both $x$ and $y$ are small. Such
contributions would not have the form of a contribution to an NRQCD
operator matrix element and would invalidate the NRQCD factorization
formula.

%\subsection{Results}

Combining NNA resummation with the complete NLO calculation, we obtain 
$
R^{\rm NNA}=(3.01\pm 0.30\pm 0.34)\times 10^3$, 
for $\alpha_s(2m_c)=0.247\pm 0.012$. This is about 50\% larger than 
$R^{\rm NLO}(2m_c)$ and is in good agreement with experiment.
% In good agreement with $R^{\rm Exp}=(3.3\pm 1.2)\times 10^3$.
The first uncertainty arises from the uncertainty in $\alpha_s(2m_c)$.
The second uncertainty comes from a velocity-scaling estimate of the
unknown matrix element of $\mathcal{O}_8({}^3S_1)$, which we simply
treat as an error of order $v^3$.
%$\langle\eta_c|\mathcal{O}_8({}^3S_1)|\eta_c\rangle$, which is of order
%$v^3$.
In the resummed expression, the uncertainty from the choice of $\mu$ is
reduced to about 11\% because the bubble sum is a renormalization-group
invariant (at leading order in the $\beta$ function), and, hence, only
the residual non-bubble-sum contribution depends on $\mu$. As is the
case with any resummation, the result depends on the choice of
resummation scheme. We can check that our choice is reasonable by
comparing with a different resummation scheme, namely, the use of the
background-field-gauge gluon vacuum polarization instead of naive
non-Abelianization. This scheme yields a very similar result: $R^{\rm
BFG}=(3.26\pm 0.31\pm 0.47)\times 10^3$.

%\subsection{Why Doesn't the BLM Method Resum the Bubbles?}

One may ask why the BLM method, which is supposed to account for 
vacuum-polarization corrections, yields a different result 
than NNA resummation. 
The BLM method approximately resums the vacuum-polarization correction 
through a choice of scale. A change of scale generates a geometric series:
$
\alpha_s(\mu')\approx
%\alpha_s(\mu){1\over 
%1-2\alpha_s(\mu)\beta_0\ln(\mu/\mu')}=
\alpha_S(\mu)[1+2\alpha_s(\mu)\beta_0\ln(\mu/\mu')
+\ldots]$.
However, in the NNA resummation, the geometric series in the integrand
may not yield a geometric series after integration over the gluon
virtualities. For example,
$
G_1(2m_c)=
1+1.91 \alpha_s(2m_c) +2.47 \alpha_s^2(2m_c) +0.97\alpha_s^3(2m_c)
-4.49\alpha_s^4(2m_c)
-11.76\alpha_s^5(2m_c)+\cdots$, which is nothing like a geometric 
series.

%\subsection{Comments}

\section{Summary and Discussion}

%The NRQCD factorization approach provides a systematic method for
%calculating quarkonium decay production rates as a double expansion in
%powers of $\alpha_s$ and $v$. 
The NRQCD factorization approach has had
significant successes in predicting the rates for a number of processes.
Among these are inclusive $P$-wave quarkonium decays, quarkonium
production at the Tevatron, $\gamma\gamma\rightarrow J/\psi +X$ at LEP,
and quarkonium production in deep-inelastic scattering at HERA. Other
processes are, so far, more problematic. The predictions of NRQCD
factorization are unconfirmed for quarkonium polarization at the
Tevatron, inelastic quarkonium photoproduction at HERA, and  double
$c\bar c$ production at Belle. For the former two processes, 
owing to large error bars, there is no real discrepancy at present. 

%It
%should be noted that, in the case of production rates, one is testing
%not only the validity of NRQCD effective field theory, but also the
%validity of hard-scattering factorization. Corrections to
%hard-scattering factorization are thought to be suppressed by powers of
%$mv/p_T$.

More precise theoretical predictions are hampered by uncertainties in
the NRQCD matrix elements and large corrections in NLO in $\alpha_s$ and
$v$, which have cast some doubt on the convergence of the series in 
$\alpha_s$ and $v$. Lattice calculations can help to pin down the decay
matrix elements. It is not yet known how to formulate the
calculation of production matrix elements on the lattice. Calculations
of higher order in $v$ are still lacking for many processes, but there is
now hope that the $v$ expansion settles down in order $v^4$. In
general, resummation of large corrections to the $\alpha_s$ series are
needed. Standard methods exist for dealing with logarithmic corrections.
The bubble resummation method is a promising technique for dealing with
nonlogarithmic corrections associated with vacuum-polarization
contributions. However, other types of large nonlogarithmic corrections are
not yet understood. A possible clue to their analysis may lie in the fact
that large order-$\alpha_s$ corrections seem to be correlated with large
order-$v^2$ corrections. 

It is clear that there are still many
interesting and challenging problems in heavy-quarkonium physics that
remain to be solved.

\subsection{Acknowledgements}

This work in the High Energy Physics Division at Argonne National
Laboratory is supported by the U.S.\ Department of Energy, Division of
High Energy Physics, under Contract W-31-109-ENG-38.

\end{document}